# Fluctuation-Enhanced Sensing: Review of Practical Issues

Graziella Scandurra,[1] Janusz Smulko,[2] and Laszlo B. Kish[3]

[1] Department of Engineering, University of Messina, Messina I-98166, Italy.
[2] Department of Metrology and Optoelectronics, Gdańsk University of Technology, Gdańsk 80-233, Poland.
[3] Department of Electrical and Computer Engineering, Texas A&M University, College Station, TX 77843-3128, USA

Correspondence should be addressed to Graziella Scandurra; gscandurra@unime.it

## Abstract

We discuss some of the fundamental practical limitations of the Fluctuation-Enhanced Sensing of odors and gases. We address resolution, measurement speed, reproducibility, memory and other problems such as humidity. Various techniques and ideas are presented to overcome these problems. Circuit solutions are also discussed.

## 1. Introduction: What is Fluctuation-Enhanced Sensing [1]

The conception and development of the Fluctuation-Enhanced Sensing (FES) method indicates that scientists in the field of sensing have realized that the microscopic, random fluctuation phenomena in physical systems are rich sources of information.

The classical way of physical and chemical sensing involves the measurement of the value of a physical quantity in the detector/sensor. FES was first proposed and developed for chemical and gas sensing that mimics the biological way of sensing in the analogy that the sensed agent changes the statistics of the neural output, which is a pulse noise. Thus, noise carries the sensory information. Note that FES has been used for a long time to measure certain physical quantities under difficult conditions. For example, the measurement of Johnson noise voltage of resistors has been utilized to determine temperature in cryogenic applications.

In Figure 1, the usual Johnson voltage noise thermometry is compared with classical resistor thermometry. For a resistor thermometer, the $R(T)$ function must be known and a biasing current that is heating the thermometer, thus it causes errors, is used to measure the actual $R$ value. In the usual Johnson-noise thermometry, the $R(T)$ function is not needed but the resistance value $R$ still must be measured thus the same problem exist, unless its value is independent of the temperature.

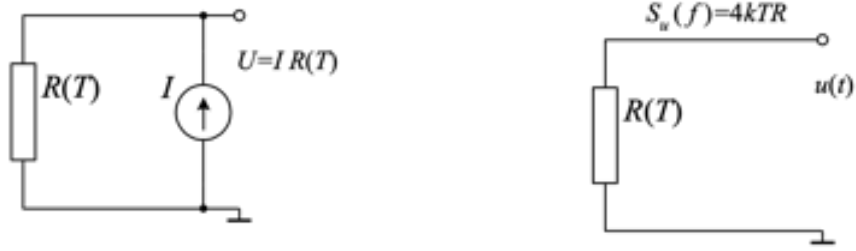

Figure 1. Left: Resistor thermometer; the $R(T)$ function must be known and a biasing DC current that is heating the thermometer and causes error. Right: The usual Johnson-noise thermometry: $S_U$ is the power spectral density of the voltage fluctuations across the resistor, $k$ is the Boltzmann constant and $T$ is the absolute temperature. The $R(T)$ function is not needed but the resistance value $R$ must be known.

In Figure 2, the method utilizes the full power of FES. This absolute thermometry is based on Johnson noise and first principles. The Johnson voltage noise and current noise are measured. They result in a system of two separate equations. Their solution provides both the absolute temperature $T$ and the resistance $R$:

$$T = \frac{\sqrt{S_u S_i}}{4k} \qquad R = \sqrt{\frac{S_u}{S_i}} \qquad (1)$$

In equation (1) $S_u$ and $S_i$ are the Power Spectral Densities (PSD) of the voltage and current fluctuations, respectively and $k$ is the Boltzmann constant. No heating occurs and only precise current and voltage measurements are needed. This solution also indicates that FES is a powerful tool, it does not mean that it is an easy/cheap measurement.

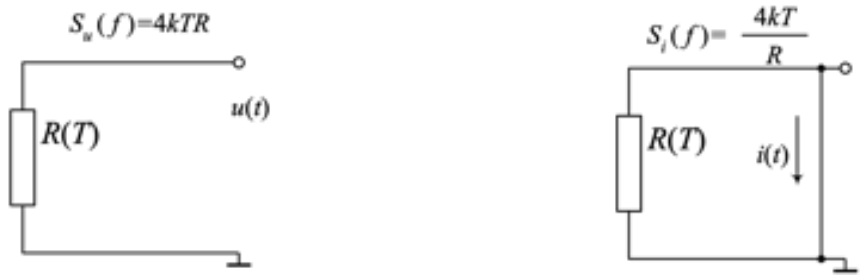

Figure 2. Fluctuation-enhanced absolute thermometry based on Johnson noise and first principles. Left: The Johnson voltage noise is measured. Right: The Johnson current noise is measured. The combination of the two measurements provides both the absolute temperature $T$ and the resistance $R$. No heating occurs and no preliminary knowledge is needed except precise current and voltage measurements.

The focus topic of our paper is *fluctuation-enhanced sensing of odors,* which is far less "clean" than the first principle measurement mentioned above. Classical gas sensing methods are many orders of magnitude less sensitive than the nose of dogs or even that of humans. So, how do biological noses do the job? As it is shown in Figure 3, they contain a large array of olfactory neurons that communicate stochastic voltage spikes to the brain. When odor molecules are adsorbed by a number of neurons, the statistical properties of these stochastic spikes change. The brain decodes the changes in statistics and matches the result with an odor database in memory.

The first step is similar to the classical sensing method: the value of a physical quantity in the sensing medium is measured, for example the output voltage of a chemical sensor. Then the microscopic spontaneous fluctuations of these measurements are strongly amplified (typically 1-100 thousand times) and the statistical properties of these fluctuations are analyzed.

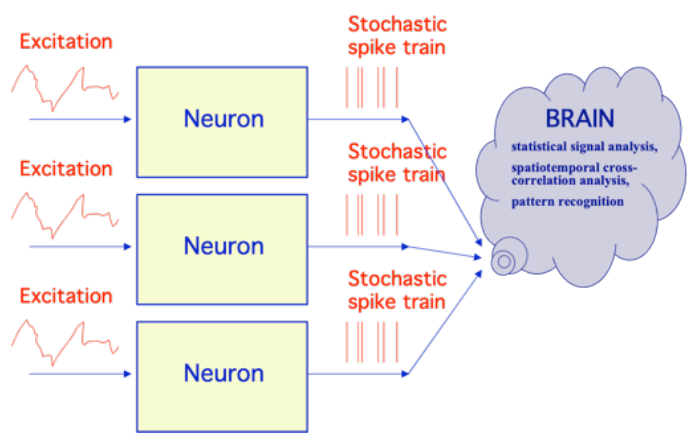

Figure 3. Fluctuation-enhanced odor and taste sensing by humans and animals.

These fluctuations are due to the dynamically changing molecular-level interactions between the odor molecules and the sensing media, thus they contain the chemical signature of the odor. The results are classified with a statistical pattern database to identify the odor [2-13]. The method has been named "fluctuation-enhanced" sensing [14-20].

In the next sections we first address the problem of classical gas sensing and then the history of FES. Then we will address some of the many practical problems of FES.

## 1.1. Short survey of the history of fluctuation-enhanced gas sensing

While some optical chemical sensors analyze the absorption or emission spectrum of gases and, therefore, they are able to generate a pattern, most chemical sensors produce a single number output only. For example, the steady-state value of a Taguchi sensor [21], or the steady-state current value of a MOS sensor, are such signals. To generate a separate pattern corresponding to different chemical compositions, a number (6 to 40) of different types of sensors are needed, which makes the system expensive and unreliable for practical applications. On the other hand, Fluctuation-Enhanced Sensing (FES), see Figure 4, is able to generate a complex pattern by the application of a single sensor [22-30].

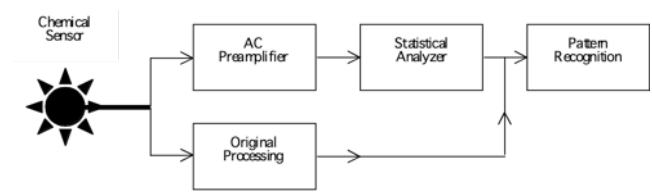

Figure 4. Example for fluctuation-enhanced chemical sensing.

FES means that, instead of using the mean value (time average) of the sensor signal, the small stochastic fluctuations around the mean value are amplified and statistically analyzed. Due to the grainy structure of resistive film sensors, these materials exhibit significantly (several orders of magnitude) increased electronic resistance fluctuations compared the case of single crystalline materials, and these fluctuations are strongly influenced by the random walk (diffusion) dynamics of agents in the vicinity of intergrain junctions and by adsorption-desorption noise. Stochastic analytical tools are used to generate a one-dimensional or two-dimensional pattern from the time fluctuations. The analysis of these patterns can be done in the classical way by using pattern recognition tools. A typical fluctuation enhanced gas sensing setup is shown in Figure 5.

The history of FES goes back more than a decade [22-44]. The name "Fluctuation-Enhanced Sensing" was created by John Audia (SPAWAR, US Navy) in 2001.

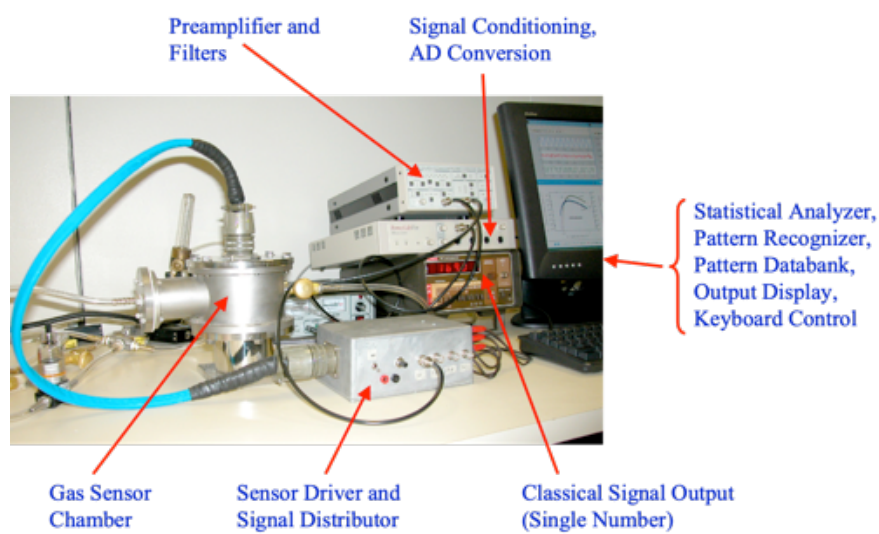

Figure 5. Example of a fluctuation-enhanced gas sensing setup.

Using electrical noise (spontaneous fluctuations) to identify chemicals was first proposed by Bruschi and coworkers [22,23] in 1994-95 by showing the sensitivity of conductance noise spectra of conducting polymers as a function of the ambient gas composition. In 1997, Gottwald and coworkers [24] published similar observations about the conductance noise spectrum of semiconductor resistors with non-passivated surfaces. The first scheme of a generic FES systems for quantitative analysis of gas mixtures with mathematical analysis about the limits and with the sensor number requirement versus the number of agents, was done by Kish and coworkers in 1998 [25-27]. The possibility of "freezing the odor", that is, the sampling-and-hold technique (see Section 4.2.1) in a Taguchi sensor was first demonstrated in [26] and later a more extensive analysis was published by Solis et al. [28]. In 2001, Smulko et al. were the first to use higher-order statistics to enhance the extracted information from the stochastic signal component [29,36,37]. Hoel et al showed FES via invasion noise effects at room-temperature in nanoparticle films [30]. Schmera and coworkers analyzed the situation of Surface Acoustic Wave (SAW) sensors and predicted the FES spectrum for SAW and MOS sensors with surface diffusion [31-32]. Commercial-Off-The-Shelf (COTS) sensors with environmental pollutants and gas combinations were also studied [33,37,38]. In nanoparticle sensors with a temperature gradient, the possibility of using the noise of the thermoelectric voltage for FES was demonstrated [36]. Ederth et al. analyzed the sensitivity enhancement in the FES mode and compared it to the classical mode in nanoparticle sensors and found an enhancement of a factor of 300 [39]. Gomri et al. [40,41] published FES theories for the cases of adsorption-desorption noise and chemisorption-induced noise. Huang et al. explored the possibility of using FES in electronic tongues [42].

## 2. Fundamental resolution limits

### 2.1 On the sensitivity and selectivity in fluctuation-enhanced sensing

The statistics of the microscopic fluctuations in a system are rich and sensitive sources of information about the system itself. They are extremely sensitive because the perturbations of microscopic fluctuations require only very small energy. On the other hand, the related statistical distribution functions are data arrays, and thus they can contain orders of magnitude more information than a single number represented by the mean value of the sensor signal used in classical sensing.

The underlying physical mechanism behind the enhanced *sensitivity* is the temporal fluctuations of the agent's or its fragment's concentration at various points of the sensor volume where the sensitivity of the resistivity against the agent is different. This effect will generate stochastic fluctuations of the resistance and the sensor voltage during biasing the sensor with a DC current. The voltage fluctuations can be extracted (by removing the mean value by AC coupling) and strongly amplified. Sensitivity enhancement by several orders of magnitude was demonstrated by Kish and coworkers [37] in Taguchi sensors and by Ederth and coworkers [39] in nanoparticle films.

Significantly increased *selectivity* can be expected depending on the type of sensor and types of available FES "fingerprints". We define the selectivity enhancement by the factor specifying how many classical sensors a fluctuation-enhanced sensor can replace. When using power density spectra, the theoretical upper limit of selectivity enhancement is equal to the number of spectral lines. At typical experiments that is about 10000. However, when the elementary fluctuations are random-telegraph signals (RTSs) the underlying elementary spectra are Lorentzians [43,44] and the situation is less favorable because their spectra strongly overlap. As a consequence, experiments with COTS sensors indicate that the response of spectral lines against agent variations is often not independent. In a simple experimental demonstration with COTS sensors, a selectivity enhancement of six was easily reachable [18]. However, nanosensor development may be able to use all of the spectral lines more independently. Because both the FES signal in macroscopic sensors and the natural conductance fluctuations of the resistive sensors usually show $1/f$-like spectra [43,44], the lower the inherent $1/f$ noise strength in the sensor the cleaner the sensory signal. An interesting analysis can be made if we suppose that we shrink the sensor size so much that the different agents probe different RTS signals. Then principles for $1/f$ noise generation [45] indicate that one can resolve at most a few Lorentzian components in a frequency decade. Supposing six decades of frequency, the maximal selectivity enhancement would be around 18, supposing three fluctuators/decade.

With bispectra [29,34,35], the potential of selectivity increase is even greater because bispectra are two-dimensional data arrays. In the case of 10000 spectral lines, as mentioned above, the theoretical upper limit of selectivity increase is 100 million, but in the Lorentzian fluctuator limit that number is again radically reduced. Bispectra sense only the non-Gaussian part of the sensor signal, and for the utilization of the full advantages of bispectra it seems necessary to build the sensor within the submicron characteristic size range in order to utilize elementary microscopic switching events as non-Gaussian components. Moreover, the sixfold symmetry of the bispectrum function yields a further reduction of information by roughly a factor six. Using the above-mentioned estimation with three Lorentzian fluctuators/decade, over six decades of frequency, the selectivity enhancement would be around 50. It should be noted that this enhancement is independent from the spectral enhancement discussed above because bispectra probe the non-Gaussian components [43].

## 2.2 Information channel capacity in classical (deterministic) gas sensors

Using Shannon's formula of information channel capacity in analog channels, it has recently been shown [43] that, in the case when the probing current density in the sensor is homogeneous and the sensor resistance fluctuations in the reference gas have $1/f$ spectrum, classical resistive sensors have the following upper limit of information flow rate:

$$C \approx \frac{1}{2t_m} ln\left[1 + \frac{8\pi^2 V(R-R_0)^2}{AR^2}\right] = \frac{1}{2t_m} ln\left[1 + \frac{8\pi^2 A_S d(R-R_0)^2}{AR^2}\right] \quad \text{(bit/s)} \quad (2)$$

Where $t_m$ is the measurement time window, $R$ and $R_0$ are the resistance response in the test gas and in the reference gas, respectively. $V$ is the volume of the sensor film, $A_S$ is the surface of the sensor film and $d$ is its thickness. The factor $A$ characterizes the strength of $1/f$ noise of the conductance noise spectrum [43] in the specific sensor by the simplified Hooge formula:

$$\frac{S_u(f)}{U^2} = \frac{A}{Vf} \quad (3)$$

where $U$ is the DC voltage drop across the resistive sensor, $f$ is the frequency and $A$ is the factor describing the strength of $1/f$ noise.

According to Equation (2), in the practical ($1/f$-noise-dominated) limit and at fixed measurement time and film thickness, the larger the surface of the classical resistive sensor the greater the information channel capacity. However, in the limit of a sufficiently large agent concentration, the saturation time is controlled by the underlying diffusion processes taking place through the thickness of the film; therefore, in this case, the shortest measurement time is also controlled by diffusion according to

$$t_{m,min} \approx \left(\frac{d}{D}\right)^2 \quad (5)$$

where $D$ is the diffusion coefficient of the agent and/or its fragments through the film. *Therefore, the thinner and larger the film the greater the information channel capacity.* This fact indicates that, in classical films, small thickness and large surface are preferable. From these two, the thickness is the dominant control parameter.

## 2.3 Information channel capacity in fluctuation-enhanced (stochastic) sensing

Current understanding of related fluctuations imply that the spectrum $S(f)$ is the superposition of elementary Lorentzian spectra $s_j(f)$. Physically such a Lorentzian fluctuator has an underlying exponential relaxation process with a single time constant $\tau_j$ determined by microphysical parameters:

$$s_j(f) = \frac{c_j}{1+2\pi f \tau_j} \quad , S(f) = \sum_j s_j(f) \tag{6}$$

where $s_j(f)$ is the spectrum originating from the $j$-th fluctuator, $c_j$ is its strength and $\tau_j$ is its time constant.

Suppose the power density spectrum of the resistance fluctuations in a FES sensor has $K$ different frequency ranges, in which the dependence of the response on the concentration of the chemical species is different from the response in the other ranges, one can write [24-26]:

$$\Delta S(f_1) = A_{1,1}C_1 + A_{1,2}C_2 + \cdots + A_{1,N}C_N$$
$$\vdots \tag{6}$$
$$\Delta S(f_K) = A_{K,1}C_1 + A_{K,2}C_2 + \cdots + A_{K,N}C_N$$

where $\Delta S(f_i)$ is the change of the power density spectrum of resistance fluctuations at the $i$-th characteristic frequency band, and the $A_{i,j}$ quantities are calibration constants in the linear response limit. Thus, a single sensor is able to provide a set of independent equations to determine the gas composition around the sensor. The number $K$ of different applicable frequency ranges has to be greater than or equal to the number $N$ of chemical species, i.e., $K \geq N$. This idealized situation needs sensors with linear response. Taguchi sensors do not produce linear response (see Figure 6) however future nanoscale devices may provide this property by the linear superposition of elementary fluctuations.

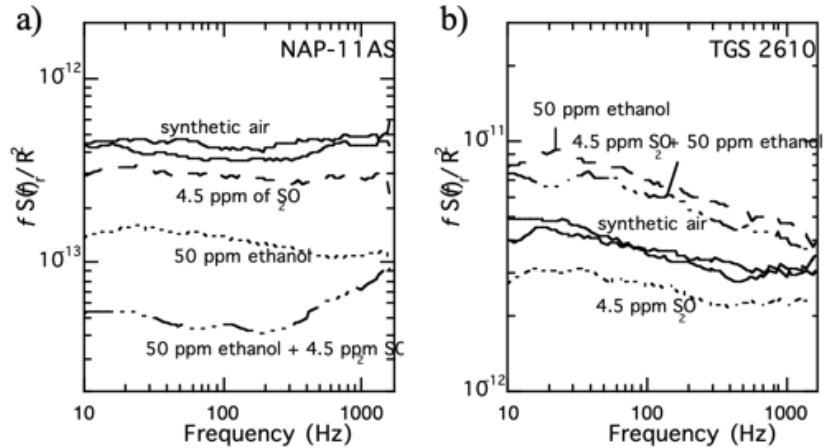

Figure 6. Experimental proof that the response in Taguchi sensors in not additive (nonlinear) [30].

In [43], the information channel capacity of power density spectrum based FES was estimated by assuming equal frequency bands and supposing that the relative error of the spectrum is much less than unity, then the information channel capacity (roughly) scales as:

$$C \propto \frac{t_m t_w f_s^2}{\Delta f} \tag{7}$$

where $t_w$ is the duration (time window) of a single data sequence (for a single Fourier transformation), $t_m$ is the total measurement time (the elementary power spectra are averaged over that) and $f_s$ is the sampling frequency. It is supposed that the FES measurement starts after the sensor reached a stationary state in the test gas and that $t_m$ is much longer than the time needed to reach the stationary state, and that condition supposes thin sensor film just like in the classical sensor considerations above.

In such a case, the most important conclusion of Equation (6) is that, in resistive FES applications, *the sensor surface can be very small without limiting the performance* (as large as enough Lorentzian fluctuators are present for each frequency band) because measurement time related statistical inaccuracies limit the information channel capacity, not the background noise.

In [43], similar conclusions are obtained for bispectrum based FES sensors.

## 3. Practical sources of errors

### 3.1 Turbulence and convection

The flow of the carrier gas around the sensor can cause vortex/turbulence effects that are random. Similarly, the convection of the hot air flying up from the heated sensors has random features. These contribute to the temperature fluctuations of heated sensors, which implies related resistance fluctuations because their resistance is strongly temperature dependent as they are typically semiconductors. These excess fluctuations are a nuisance because disturb the FES signals which are also fluctuations.

There is a method of operation, the "sampling-and-hold" (or "frozen-smell") technique, that is free of most the above-mentioned problems (see details in Section 4.2.1 below).

### 3.2 Ambient air, unknown agents and humidity

The FES spectral patterns of various gas compositions are learned by a classifier (pattern recognizer such as neural network) that will display the result of the analysis. However strictly speaking this works only for agents used in the training of the classifier, typically in a lab environment. When the FES system is used on the field, extra agents can be present. If they influence the FES spectra that poses a problem. The human nose therefore is using not only FES (via random neural signals) but also a large number of different types of olfactory neurons. Possibly, future chip technology can help with this problem by imitating biological smelling.

Humidity can also be different at the field measurements compared to lab conditions and it can have major influence on the spectra. In this case humidity must also be treated as an agent.

### 3.3 Memory, aging, fabrication variations

Noise is sensitive to the spatial correlations and their time variations in the molecular lattice of the sensor. This is the reason why FES is a much more sensitive and information rich tool than classical sensing based on resistivity changes. However, these enhancements come with a price: the noise may be too sensitive thus it may reflect on also memory, aging and variations in fabrication. All these facts must be taken into the account and carefully monitored during applications.

## 3.4 Measurement circuitry problems: noise, bandwidth

Performing sensible low frequency noise measurements (LFNM) is never an easy task. The very instrumentation required for performing noise measurements introduce noise that may mask the noise coming from the DUT, thus reducing the sensitivity that can be obtained. Since in the field of FES we are mainly interested in the low frequency noise generated by the DUT, it is generally the flicker component of the noise introduced by the instrumentation that sets the limit to the sensitivity that can be obtained. It must also be noted that LFNM measurement set-ups are extremely sensitive to interferences coming from the environment. Electro-Magnetic Interferences (EMI) at frequencies much higher than those one is interested in may translate into what could appear as additional low frequency noise components because of rectifying effects due to the active components in the measurement chain [46]. Even moderate air movement, in the immediate proximity of active components, as caused by unrestricted convection motion due to power dissipation, may result in apparent increase in the noise. Assuming that all causes of interferences are removed by proper shielding and positioning of the components, the ultimate sensitivity is set by the intrinsic noise introduced by the DUT bias systems, by the preamplifiers that detect and amplify the noise signal across the DUT and, possibly, by other auxiliary systems such as DUT temperature control systems. While standard instrumentation exists that can be used for laboratory set ups and even wafer level noise testing [47-49], depending on the DUT characteristics, dedicated instrumentation may be a better choice in order to maximize sensitivity [50] and, in the specific case of FES applications, portability. The main circuitry solutions will be discussed in section 4.3.

## 4. Reducing errors and enhancing the sensory information channel capacity

### 4.1 Photonic excitation

Conductance noise spectra are the superposition of over-damped Lorentzian spectra, which means no sharp peaks in the FES spectrum can be present. However, photonic excitations combined with FES measurements have the potential to increase the information content of measurements. With properly selected photon energies (wavelengths) some of the otherwise dormant states can be activated and they can produce a modified spectrum with new information.

#### 4.1.1 UV light excitation experiments and results
Resistive gas sensors have limited selectivity and sensitivity, and therefore any cost-efficient methods of gas sensing improvement are appreciated. Some gas sensing materials (e.g., $WO_3$, $TiO_2$, $SnO_2$ nanowires, golden nanoparticle organically functionalized, carbon nanotubes, graphene) exhibit a photocatalytic effect. This effect can be easily induced by UV LEDs irradiating the gas sensing layer (Figure 7). Numerous papers are presenting the detailed results of gas sensing modulated by UV light applied to different gas sensing layers [51-58].

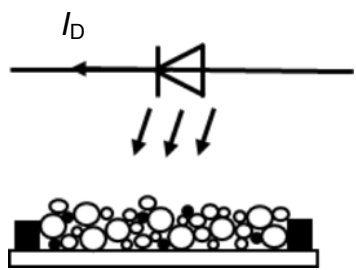

Figure 7: Illustration of resistive gas sensing layer between the terminals (black rectangles) irradiated by UV LED biased by DC $I_D$; the layer comprises of the grains of different size and dopants of noble metals (black dots) improving gas selectivity.

A series of low-cost UV LEDs, emitting UV light within wavelet range 250÷355 nm is available on the market. Resistive gas sensing layer can be made on ceramic or silicon substrates and efficiently irradiated from different distances. Selected wavelengths of the UV light can modulate the physical properties of the gas sensing layer (e.g., DC resistance, low-frequency resistance fluctuations) [55]. We observed that changes of DC resistance induced by UV irradiation are independent of the changes of the recorded low-frequency noise [56]. Thus, both phenomena secure additional information about the ambient atmosphere and can be utilized to improve gas detection.

The FES method considers power spectral densities of resistance noise for enhanced gas detection. It is a function of frequency and can deliver more information than a single DC resistance value. We observed 1/f dependence with a plateau at selected corner frequencies, characteristic for an ambient gas and the wavelength of the applied UV LEDs. The intensity of the 1/f noise component was also informative and depended on gas concentration. The FES method was advantageous at low gas concentrations enhancing threshold gas detection. This effect was observed at an ambient atmosphere of different gases. It is a valuable result for practical applications, especially for medical diagnosis by exhaled breath analysis when deficient gas concentrations have to be detected. Moreover, we observed faster response of the gas sensor. This effect is rather evident because UV light was formerly applied to cleanse the sensor. UV irradiation replaced a heating pulse, commonly employed for fast sensor cleansing.

We should underline that a photocatalytic effect in gas sensing depends on a few independent factors:
- morphology of gas sensing layer, determining the depth of light penetration,
- irradiation intensity,
- the wavelengths of the UV light.

As we see (Figure 7), the gas sensing layer is a porous grainy material. The gas molecules can penetrate the pores of the whole gas sensing layer due to the diffusion process and change DC resistance or low-frequency resistance noise recorded between the terminals. The UV light penetrates only the thin part of the gas sensing layer. The depth of penetration depends on light wavelength, irradiated material and its morphology. We can assume that the depth of UV light penetration does not exceed a few μm only [55]. Thus, the gas sensing layer can be modelled as two parallel resistors: an external one modulated by UV light and the inner one without this effect. We observed that there is a threshold intensity of UV light when its further intensification does not induce the changes in the gas sensing layer. This effect is a result of a limited impact on the thin sensing layer only. The low-cost gas sensor in Figure 8, made of a mixture of graphene flakes (Graphene Supermarket UHC-NPD-100ML) and TiO$_2$ nanoparticles (AEROXIDE® TiO$_2$ P25) by painting and baking, has a thickness of about 100 μm, which is about hundred times more than the estimated depth of UV light penetration. Thus, UV light affects less than 1% of its porous and gas-sensitive volume.

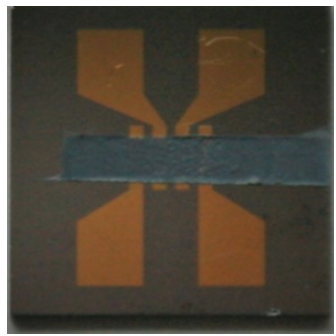

Figure 8: Gas sensing layer comprising $TiO_2$ nanoparticles and graphene flakes: photo of the sensor made by painting on a silicon substrate with four-point gold contacts.

The easiest way of increasing UV light modulation is to apply a thinner gas sensing layer. Moreover, a thinner layer will secure a faster gas response, which is very important for any practical application. The reduced thickness should result in other advantages, such as a lower price and lower energy consumption when operating at elevated temperatures. This result can be reached by spin coating technology, reducing the thickness of the sensing layer to the maximum of a few μm only and ensuring more even surface and more repeatable sensing properties.

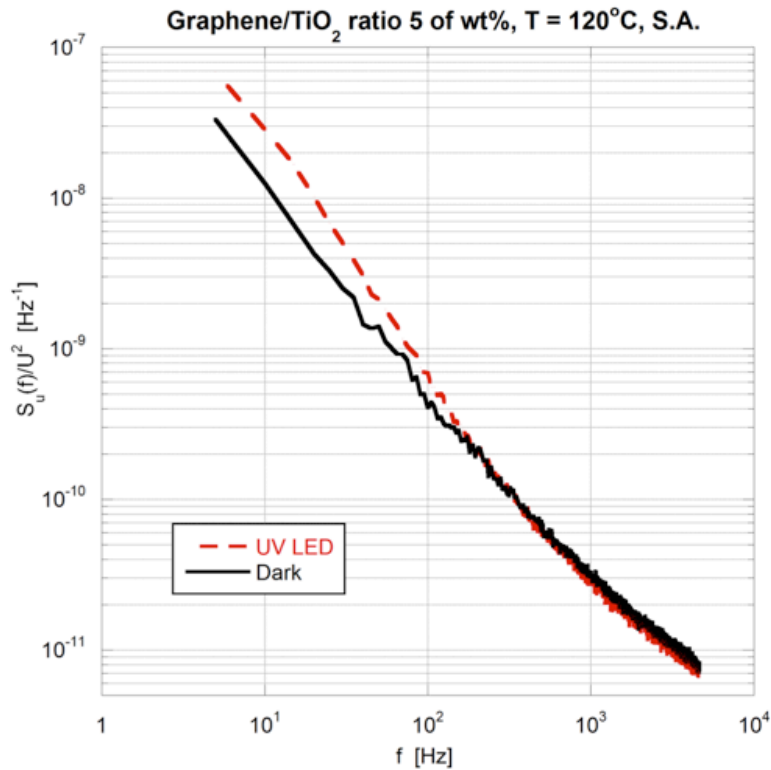

Figure 9: Power spectral density $S_u(f)$ of voltage fluctuations across the gas sensor biased by DC voltage $U$ and normalized to $U^2$ versus frequency $f$; UV LED, type OSV4YL5451B, made by OptoSupply was irradiating the sensing layer with a graphene flakes/TiO$_2$ ratio 5 of wt% (Figure 8) in the ambient atmosphere of synthetic air (S.A.) and at operating temperature $T = 120$ °C.

Figure 9 presents results of 1/f noise power spectral density generated in the thick gas sensing layer shown at Figure 8. The sensor was irradiated by UV LED, having maximal optical power at the wavelength of 394 nm. Both materials (graphene flakes, TiO$_2$ nanoparticles) are gas-sensitive and both exhibit the photocatalytic effect. The gas sensing layer has a very porous structure. We observed a vivid change of power spectral density slope versus frequency when UV light was applied. The intensified plateau of power spectral density was present about 10 Hz after irradiation. This effect can be explained by modulation of adsorption-desorption rate induced by UV light, making these processes faster and more intense. We have the possibility of strengthening this effect by increasing power of UV light (decreasing its wavelength), changing morphology of gas sensing layer (e.g., changing a graphene flakes/TiO$_2$ ratio of wt%) and reducing a thickness of the gas sensing layer. More visible changes of plateau were observed at lower operating temperatures when energy of UV light had greater impact on adsorption-desorption process.

The FES method can be applied not only in resistive gas sensing layers. The back-gated FET gas sensors, utilizing graphene in a channel, exhibit 1/f spectra with a plateau at corner frequencies characteristic for selected ambient gases (e.g., chloroform, acetonitrile, methanol, ethanol) [57]. This result was observed when a single-layer-graphene was used on the channel. Then, low-frequency noise was related to individual events of gas molecules adsorption-desorption. A corner frequency of the observed plateau was stable for different specimens of the single-layer-graphene FET sensors at an ambient atmosphere of acetonitrile [57]. This result suggests that the FES method can be very selective and repeatable when the FET channel is a two-dimensional material of the same physical structure.

Another two-dimensional material – $MoS_2$ was applied successfully for gas sensing using the FES method in back-gated FET sensors [59]. These experiments require further in-depth studies to confirm the result for other gases and other materials.

We may suppose that in two-dimensional material the adsorbed molecules change the surface potential. Therefore, the observed low-frequency noise is modified by adsorption event more similarly for a given gas molecule than in a case of the porous gas sensing layer. In a bulk polycrystalline sensor, gas molecules modulate the potential barrier between the crystal grains, which have some size distribution. Any potential barrier modulation should depend on the size of the grains adsorbing the molecules. Thus, any 1/f noise change induced by the adsorbed molecule should be less characteristic in porous gas sensors, comprised of different grains, than in the FET sensor using two-dimensional material of repeatable structure in a channel. Moreover, we can suppose that the FES method in FET sensors can determine components of gas mixtures when each plateau in 1/f noise is related to the presence of different gas molecules. This conclusion is valuable for practical application when we have to consider gas mixtures and the effects of crossing gases.

Some experimental data noticed that extended exposure to UV irradiation of 280 nm LEDs damaged graphene in a channel of the FET sensor and altered the device characteristics [60]. We suppose that too-short wavelength, i.e. too-energetic, UV light can deteriorate the graphene structure. We have not observed a similar effect of deterioration induced by UV light for thick graphene flakes/$TiO_2$ gas sensing layers irradiated by UV light of longer wavelengths (362 nm or 394 nm). However, this issue requires further and more detailed studies. The abundance of commercial UV LEDs emitting light of various wavelengths should enable future thorough investigation.

The presented exemplary experimental data of the FES method modulated by UV irradiation suggest that there is a strong potential of improving selectivity and sensitivity of gas sensing by this method, mainly when two-dimensional materials are applied. We hope to observe impressing progress in the FES method by using two-dimensional materials, either in back-gated FET sensors or in low-cost gas sensing layers.

## 4.2. Temperature variation techniques

Instead of photons the dormant states can often be activated by proper temperature changes provided that the temperature change required for the given activation energy is feasible. It should also be noted here that Dutta-Horn analysis on some Taguchi sensors showed that if the measured fluctuations were thermally activated, the potential barriers were also strongly temperature dependent [30]

### 4.2.1 Sampling-and-hold measurement technique

To get around this difficulty, sampling-and-hold ("frozen-smell") FES was introduced for heated sensors [28,61] resulting in not only higher reproducibility but also higher specificity [62,63]. During this measurement, the sensor is first heated for a short time to let the agent diffuse into the molecular lattice of the sensor film. Then the heating is switched off and the FES measurements are done on the cold sensor. The agent does not even have to be present in the ambient atmosphere because the information is stored in the molecular lattice.

### 4.3 Dedicated Instrumentation

In order to overcome the problems evidenced in section 3.4, dedicated instrumentation may be the optimal choice. To discuss the options that are available for the design of dedicate instrumentation, it can be useful to divide the possible measurement set-ups into two main categories, as quite different issues must be addressed in the two cases: Voltage Noise Measurements (VNM) and Current Noise Measurements (CNM). A block diagram of basic set ups for VNM and CNM in the case of a two terminals device (DUT) are shown in Figure

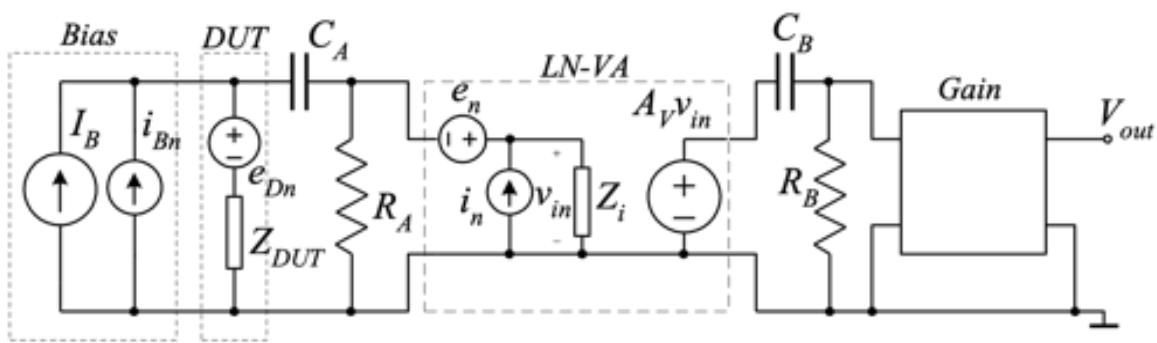

Figure 10: block diagram of basic set up for voltage noise measurements.

10 and Figure 11, respectively. In the case of VNMs, the DUT is supplied with a constant current. If the input impedance of the Low Noise Voltage Amplifier (LN-VA) is sufficiently large and its Equivalent Input Current Noise (EICN) is sufficiently low, the DC component due to the bias can be removed before the input of the LN-VA that can be therefore DC coupled and, at the same time, characterized by high gain [64-66]. Typical gains for the LNVA stage are in the order of 40 dB. A second AC filter can be used to remove the offset at the output of the LNVA prior to the input to a second stage that acts as a gain stage to insure a good coverage of the dynamic range of the signal sampler used to record and analyse the output signal. Because of the large gain of the first stage, the gain stage can often be realized employing low noise Operational Amplifier (OA) based amplifiers. On the other hand, the Equivalent Input Voltage Noise (EIVN) at the input of the LN-VA amplifier sets the ultimate level to the Background Noise (BN) that can be reached and, therefore, it is often obtained by resorting to discrete large area JFETs as the first amplifying device. Regardless of the particular architecture chosen for the realization of the LN-VA, the fact that it is based on discrete devices means that the output offset can be quite large, hence the need for a second AC coupling stage between LN-VA and the second stage [64-66]. When dimensioning the first AC coupling stage, attention must be paid to the fact that the resistance $R_A$ introduces noise that may increase the BN of the system. Assuming that $R_A$ can be made much larger than the DUT impedance, the thermal noise generated by $R_A$ is filtered by the coupling capacitance $C_A$. In order to take full advantage of the low noise of the first stage the noise contribution coming from $R_A$ must become negligible at the minimum frequency of interest $f_{min}$, and this means, as it is discussed in [67,68], that the frequency corner of the AC filter must be much smaller than $f_{min}$. The resulting long time constant may cause transients that are very long (in the order of minutes) if $f_{min}$ is in the hundred mHz range. Assuming the EICN of the LN-VA to be negligible and its input impedance to be extremely large, at frequencies at which the presence of the AC coupling filter can be neglected (both in terms of frequency response and noise contribution from $R_A$), we have that the voltage at the input of the LN-VA can be written as:

$$V_{in} = i_{Bn}Z_{DUT} + e_{Dn} + e_n \tag{8}$$

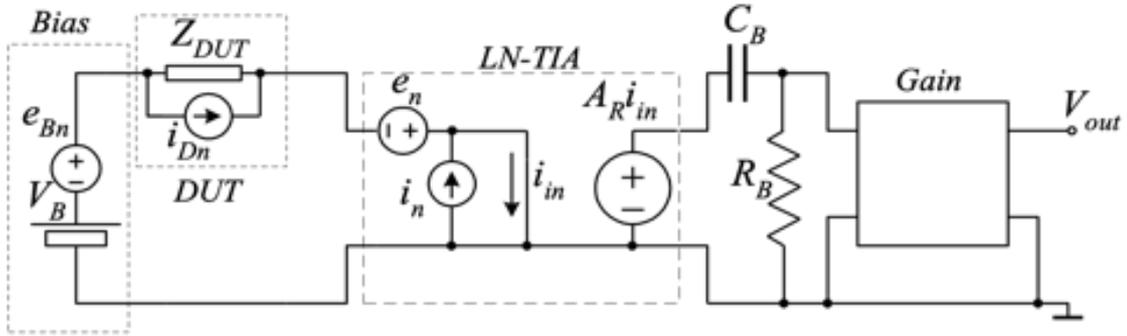

Figure 11: block diagram of basic set up for current noise measurements.

Assuming now, as it is often the case, that the noise introduced by the second stage is negligible, because of the large gain of the first stage, we can interpret the Power Spectral Density (PSD) of the noise at the output of the amplifier as due to an equivalent input voltage PSD $S_{Vin}$ given by:

$$S_{Vin} = S_{IBN}|Z_{DUT}|^2 + S_{VDN} + S_{VN} \tag{9}$$

where $S_{IBN}$ is the PSD of the noise source $i_{Bn}$, $S_{VDN}$ is the PSD of the noise source $e_{Dn}$ (i.e. the noise source generated by the DUT) and $S_{VN}$ is the PSD of the equivalent input noise source of the LN-VA.

Clearly, $S_{Vin}$ can be assumed to represent $S_{VDN}$ only if $S_{VN}$ and the effect of $S_{IBN}$ are negligible. As far as $S_{IBN}$ is concerned, a few designs for the realization of low noise current sources have been proposed [69-72]. Depending on the current range to be sourced, the designs may vary considerably and, actually, there is no single approach than can be considered sufficiently general to be regarded as a reference design for the implementation of high sensitivity, low frequency voltage noise measurement systems. In most cases, if fine tuning of the current is not required, the current source in Figure 10 is obtained starting from a battery with a resistance in series that is much larger than the DUT impedance as in Figure 12.

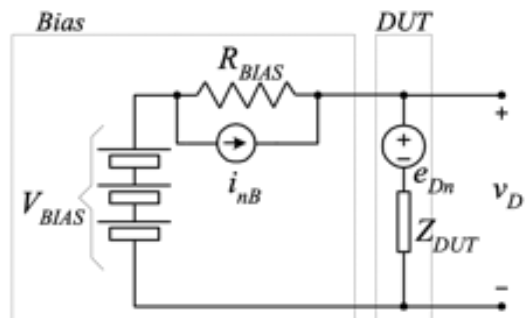

Figure 12: block diagram for the calculation of the contribution to the voltage noise across the DUT as due to the bias network

If the current supplied by the batteries ($V_{BIAS}$) is a small fraction of their maximum rated value, they behave as very low noise voltage sources, and if the resistance in series ($R_{BIAS}$) is a high quality metallic film resistance, its contribution can be reduced to its thermal noise. If, for the sake of simplicity, we assume the DUT to behave as a resistance with value $R_D \ll R_{BIAS}$, from Figure 12 we can calculate that the contribution to the voltage noise across the DUT as due to the bias network is:

$$S_{IBN} \approx \frac{4kT}{R_{BIAS}}; \quad S_{IBN}|Z_{DUT}|^2 = 4kTR_D \frac{R_D}{R_{BIAS}} = S_{VTH}(R_D) \frac{R_D}{R_{BIAS}} \tag{10}$$

where $S_{VTH}(R_D)$ is the thermal noise of the DUT. Equation (10) is particularly interesting when we observe that in order to detect low frequency noise generated by the DUT, its level must necessarily be above the thermal noise of the DUT.

Therefore, since the noise contribution of the bias network is always below the thermal noise of the DUT ($R_D \ll R_{BIAS}$), in the bias configuration in Figure 12 the noise introduced by the bias network is always negligible. Note that the condition $R_D \ll R_{BIAS}$ is also required since, in Figure 12, the contribution of the noise coming from the DUT is attenuated because of the presence of a finite $R_{BIAS}$. However, as long as $R_{BIAS} \gg R_D$, the attenuation is small and can be usually neglected.

As we have noted above, in order to obtain very low level of BN at very low frequencies ($f$<1 Hz) we need to resort to discrete devices for the front end of the LN-VA. In order to understand the advantage that can be obtained in terms of BN when resorting to discrete JFETs, it is sufficient to observe that if we resorted to the conventional OA based voltage amplifier topology for the realization of the LN-VA, the equivalent input voltage noise source $e_n$ in Figure 10 could be made to essentially coincide with the equivalent input voltage noise of the OA. The JFET input operational amplifiers with the lowest level of equivalent input voltage noise that we have been able to find are the ones belonging to the OPAx140 series by Texas Instruments [73]. Their equivalent input voltage noise is about 50 nV/√Hz at 100 mHz, less than 16 nV√Hz at 1 Hz and about 5 nV√Hz for $f$>1 kHz. If, on the other hand, we design the LN-VA using the very large area IF3601 by InterFet [74], the equivalent input voltage noise can be as low as 5.6 nV/√Hz at 100 mHz and 1.4 nV/√Hz [65], with a gain, in terms of BN PSD reduction of 80 and 13 at 100 mHz and 1 Hz respectively. While obtaining the ultimate noise performances in discrete JFET based amplifiers may require resorting to relatively complex circuitry [64,65], optimized design aimed at simplifying implementation while maintaining excellent noise performances have been proposed. For instance, Figure 13 reports the complete schematic of the amplifier proposed in [66] that, with a very limited component count, allows to obtain an equivalent input noise of about 14, 1.4 and 0.8 nV/√Hz at 100 mH, 1 Hz and for $f$>1 kHz, respectively.

A fully differential input low noise amplifier can be derived from the circuit in Figure 13 for noise measurement across a DUT with none of the two terminals grounded [75].

Voltage noise measurements are typically employed for low to moderate impedances for the DUTs (from a few Ω up to a few tens of kΩ), where all the conditions that allow to simplify the measurement set up discussed above can be met. For much higher impedances, current noise measurements are usually more easily implemented. In principle, one could assume that current noise measurements present issues that are dual with respect to the case of voltage noise measurements, but in the case of low frequency noise measurements this is seldom the case. As we have discussed above, in the case of voltage noise measurements, AC coupling down to the tens of mHz range for removing the DC component can be obtained with simple RC networks ($R_A C_A$ in Figure 10) with reasonable values (MΩ for the resistances and tens of μF for the capacitances) with the capacitors only marginally deviating from ideality. In the case of current noise measurement, we would require an input AC coupling network capable of separating the DC bias current from the current fluctuations down to the mHz range. This could be done, in principle, by employing the circuit configuration in Figure 14 in which the $L_A R_A$ network would operate dually with respect to the $R_A C_A$ network in Figure 13. However, even assuming a resistance $R_L$ in the order of 1 kΩ (in the case of current measurements $R_L$ must be much lower with respect to the impedance of the DUT for its attenuation effect to be negligible) the value of $L_A$ that would be required to obtain a low frequency corner below 100 mHz would be larger than 1 kH, that is, to say the least, completely outside the range of values that can be realized, at least with high quality and for signal applications.

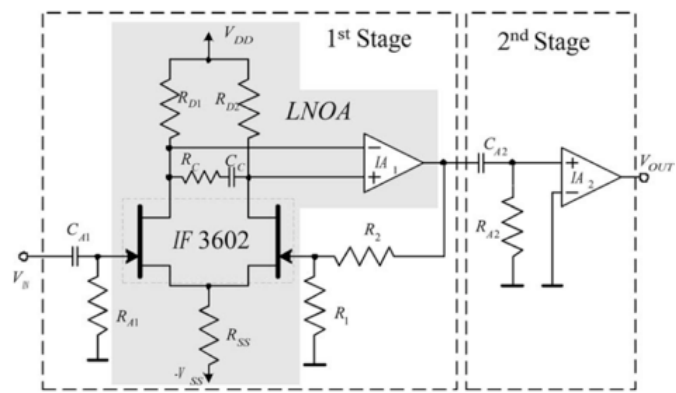

Figure 13: Complete schematic of the LN-VA proposed in [66].

This means, as it is shown in Figure 11, that the first stage (the transimpedance amplifier in Figure 11) is always coupled in DC, and this translates in an important constraint in the performances that can be obtained by a low noise transimpedance amplifier as we shall presently discuss. To this purpose, rather than dealing with the general circuit configuration in Figure 11 it is convenient to refer to the actual circuit configurations employed for the realization of Low Noise Transimpedance Amplifier (LN-TIA). In almost any instance of low frequency noise measurements, the basic circuit configuration that is employed is shown in Figure 15.

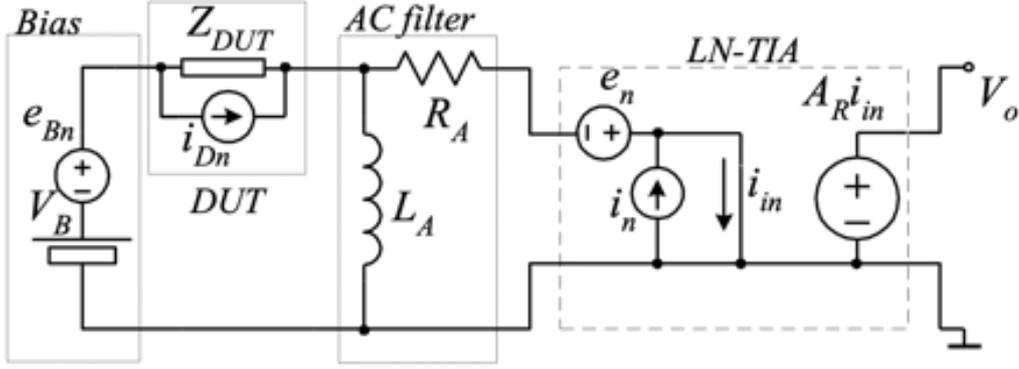

Figure 14: Ideal circuit for separating the current fluctuations entering the Low Noise – Transimpedance Amplifier (LN-TIA) from the DC bias current (ideally flowing through the inductance LA).

The operational amplifier in Figure 15 is a JFET or MOSFET input operational amplifier, so that the current noise sources that are present at its inputs can be typically neglected with respect to other sources of noise [76]. The operational amplifier equivalent input voltage noise source $e_{Vn}$ is shown explicitly in the figure as its effect can be relevant especially at higher frequencies and in the case of DUTs with large capacitive components, as it will be presently discussed. With the equivalent circuit in Figure 11 in mind, in the assumption of virtual short circuit between the inverting and non-inverting inputs of the operational amplifier, the transresistance gain $A_R$ can be calculated to be $-R_R$ while for the equivalent input voltage and current noise sources $e_n$ and $i_n$ we have:

$$i_n = -\frac{e_{Vn}}{R_R} - \frac{e_{Rn}}{R_R}$$
$$e_n = -e_{Vn}$$
(11)

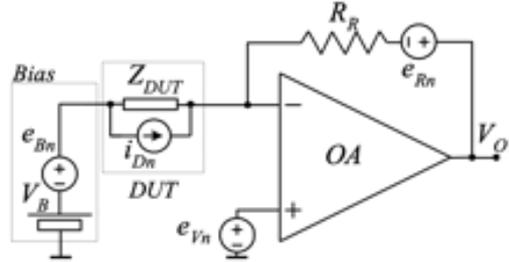

Figure 15: Reference circuit for current noise measurements.

As it is apparent from Equation (11), the equivalent sources $i_n$ and $e_n$ are correlated since $e_{Vn}$ contributes to both sources. The noise introduced by the feedback resistance ($R_R$) is due to its thermal noise and its power spectra density $S_{VR}$ is given by:

$$S_{VR} = 4kTR_R \quad (12)$$

For resistances $R_R$ in excess of 1 MΩ, the thermal noise of the resistance is much larger than $1\times10^{-14}$ V$^2$/Hz (or 100 nV/√Hz) so that, unless we need to work at very low frequencies ($f \ll 1$Hz), the contribution of $e_{Vn}$ to the equivalent input noise $i_n$ can be neglected (the PSD of $e_{Vn}$ for the TLC 2201 MOSFET input OA is in the order $3.6\times10^{-15}$ V$^2$/Hz at 1 Hz and $43.6 \times 10^{-16}$ V$^2$/Hz at 10 Hz). In this assumption we obtain, for the PSD $S_{IN}$ of the equivalent input noise source $i_n$:

$$S_{IN} \approx \frac{4kT}{R_R} \tag{13}$$

Equation (13) is particularly interesting since it shows that in order to reduce the equivalent input noise of a LN-TIA realized as in Figure 15, the feedback resistance must be made as large as possible. Unfortunately, since also the bias current flows through the feedback resistance, there is a limit at the maximum value of $R_R$ that can be used. If $I_B$ is DC bias current sourced by $V_B$, the DC output voltage $V_{ODC1}$ at the output $V_{O1}$ in Figure 15 is:

$$V_{ODC1} = -R_R I_B \tag{14}$$

For the OA in Figure 15 to remain in linearity, $V_{ODC1}$ must remain within the power supply voltages, typically a few V. Therefore, in performing current noise measurements we are in a situation where we could, in principle, obtain an equivalent input current noise as low as desired, by increasing the feedback resistance. However, increasing $R_R$ reduces proportionally the maximum current that can flow through the DUT. The fact that the DC output voltage is proportional to the bias current flowing through the DUT has however one useful outcome: by measuring the DC voltage at the output $V_{O1}$, we obtain the value of the DC current through the DUT for any given bias voltage $V_B$. The AC filter $C_B R_B$ in Figure 11 is used to remove the (large) DC component and to allow the amplification of the noise only up to a level compatible with the dynamic of the signal acquisition and elaboration system used for spectral estimation.

In order to evaluate the overall noise contribution to the output and to analyze to what extent the output noise in the circuit in Figure 15 can be regarded as essentially due to the DUT we must also take into account the effect of the noise introduced by the bias system ($e_{Bn}$) and the noise introduced by the equivalent input voltage noise $e_{Vn}$ when regarded as responsible for the equivalent input voltage noise $e_n$. The voltage noise $V_{On}$ at the output $V_{O1}$ can be obtained as:

$$V_{On} = -e_{Bn}\frac{R_R}{Z_{DUT}} + e_{Vn}\left(1 + \frac{R_R}{Z_{DUT}}\right) + e_{Rn} - i_{DN} R_R \tag{15}$$

The PSD of $V_{On}$ can be then obtained as (the noise sources in Equation (15) are uncorrelated):

$$S_{VO} = S_{VB}\left|\frac{R_R}{Z_{DUT}}\right|^2 + S_{VV}\left|1 + \frac{R_R}{Z_{DUT}}\right|^2 + 4kT R_R + S_{ID} R_R^2 \tag{16}$$

By dividing Equation (16) by the transresistance gain squared ($R_R^2$) we can interpret the noise at the output as due to the PSD $S_{IN}$ of a current source at the input of a noiseless TIA with the same gain:

$$S_{IN} = \frac{S_{VB}}{|Z_{DUT}|^2} + S_{VV}\left|\frac{1}{R_R} + \frac{1}{Z_{DUT}}\right|^2 + \frac{4kT}{R_R} + S_{ID} \tag{17}$$

Cleary we can interpret $S_{IN}$ as due to the DUT ($S_{ID}$) as long as all other contributions can be neglected. In a situation in which a sufficiently large $R_R$ can be used for reducing the contribution of its thermal noise to a negligible level, the noise introduced by the equivalent input voltage noise of the amplifier and from the bias system may become relevant. If the impedance of the DUT is large or in the same order of the feedback resistance and the noise introduced by the bias system is comparable to that due to the equivalent input voltage noise of the amplifier, the argument leading to Equation (13) can be extended to Equation (17) and, at least at low frequencies, the contributions by $S_{VV}$ and $S_{VB}$ can be assumed negligible. However, if the DUT has a significant parallel capacitive component, that can also be the result of the parasitic capacitances in the cables connecting the DUT to the system, the impedance of the DUT decreases with frequency and the noise introduced by $S_{VV}$ (and possibly by $S_{VB}$) can soon mask the noise coming from the DUT [77]. Note, moreover, that methods have been proposed that allow to reduce the noise introduced by the feedback resistance either by cross correlation approaches [78,79] or by exploiting configurations in which the feedback impedance is replaced by a capacitor that, intrinsically, does not introduce noise (noise however, is still introduced by the DC path required to close the DC feedback path from the output of the OA to the inverting input) [80-82]. In all these cases, the ultimate level of background noise is set by the equivalent input voltage noise source of the operational amplifier, assuming that the noise of the bias system can be made negligible.

As far as the bias system is concerned, in the configuration in Figure 15 it is the voltage source $V_B$ that supplies the DC current to the DUT. A different configuration can be used, however, in which the voltage source providing the voltage bias does only supply the extremely low current required to bias the non-inverting input of the OA. This circuit configuration is shown in Figure 16.

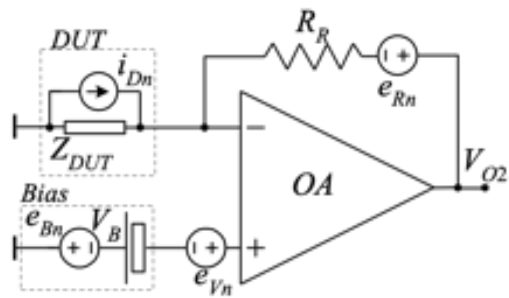

Figure 16: Alternative circuit for current noise measurements.

Because of the virtual short between the inputs of the operational amplifier, the voltage bias across the DUT in Figure 16 is the same as in Figure 15 for the same value of $V_B$. The voltage source $V_B$ must only supply the very low bias current at the input of the JFET (or MOSFET) input operational amplifier. Moreover, in this configuration the DUT has got one terminal connected to system ground. The price to be paid is the fact that in the circuit in Figure 16, the DC output voltage $V_{ODC2}$ is:

$$V_{ODC2} = -V_B - R_R I_B \qquad (18)$$

According to Equation (18), $|V_{ODC2}|$ is larger than $|V_{ODC1}|$ in Equation (14) for the same bias condition for the DUT. This results in a lower maximum bias current for remaining in linearity or, to maintain the same maximum bias current as in Figure 15, a lower value for $R_R$ and, hence, a higher level of background noise.

The circuit topologies and design guidelines summarized above can be used for the development of dedicated instrumentation tailored to the specific sensing application at hand. System integration, that is the ability to combine and optimize the electronics sections together with the mechanical structure (for support and for electromagnetic shielding) and dedicated control and data elaboration software can play a significant role in the realization of effective and user friendly noise measurement system [83-86]

## 5. Conclusions

We highlighted the problems of classic gas and odors sensing and presented, also illustrating its history, the technique called Fluctuation Enhanced Sensing as a valid and powerful alternative. The main practical problems to be addressed when applying FES have also been addressed and possible solutions have been presented.

## Data Availability

The data used to support the findings of this study are in part included within the article and available from the corresponding author upon request.

## Conflicts of Interest

The authors declare that there is no conflict of interest regarding the publication of this paper.

(1)